\documentclass[conference]{IEEEtran}

\ifCLASSINFOpdf
  \usepackage[pdftex]{graphicx}
\else
  \usepackage[dvips]{graphicx}
\fi

\usepackage{subfig}
\usepackage{graphics}
\usepackage{graphicx}

\begin{document}
\title{Special Session - Fault-Tolerant Deep Learning: A Hierarchical Perspective}

\author{
\IEEEauthorblockN{Cheng Liu, Zhen Gao, Siting Liu, Huawei Li, Xiaowei Li}
\IEEEauthorblockA{xxxx
}
}


\author{\IEEEauthorblockN{Cheng Liu \IEEEauthorrefmark{1},
Zhen Gao \IEEEauthorrefmark{2},
Siting Liu \IEEEauthorrefmark{3}\IEEEauthorrefmark{5}, 
Xuefei Ning \IEEEauthorrefmark{4},
Huawei Li \IEEEauthorrefmark{1}\IEEEauthorrefmark{6} and
Xiaowei Li \IEEEauthorrefmark{1}}
\IEEEauthorblockA{\IEEEauthorrefmark{1} SKLCA, Institute of Computing Technology, Chinese Academy of Sciences\\}
\IEEEauthorblockA{\IEEEauthorrefmark{2} School of Electrical and Information Engineering, Tianjin University\\}
\IEEEauthorblockA{\IEEEauthorrefmark{3} School of Information Science and Technology, ShanghaiTech University\\}
\IEEEauthorblockA{\IEEEauthorrefmark{4} Department of Electronic Engineering, Tsinghua University\\}
\IEEEauthorblockA{\IEEEauthorrefmark{5} Shanghai Engineering Research Center of Energy Efficient and Custom AI IC}
\IEEEauthorblockA{\IEEEauthorrefmark{6} Peng Cheng Laboratory, Shen Zhen}
}

\maketitle

\begin{abstract}
With the rapid advancements of deep learning in the past decade, it can be foreseen that deep learning will be continuously deployed in more and more safety-critical applications such as autonomous driving and robotics. In this context, reliability turns out to be critical to the deployment of deep learning in these applications and gradually becomes a first-class citizen among the major design metrics like performance and energy efficiency. Nevertheless, the back-box deep learning models combined with the diverse underlying hardware faults make resilient deep learning extremely challenging. In this special session, we conduct a comprehensive survey of fault-tolerant deep learning design approaches with a hierarchical perspective and investigate these approaches from model layer, architecture layer, circuit layer, and cross layer respectively.
\end{abstract}


\IEEEpeerreviewmaketitle
\section{Introduction}
Deep learning has been demonstrated to be successful in a plethora of applications including computer vision \cite{krizhevsky2012imagenet} \cite{redmon2016you} and natural language processing \cite{young2018recent}, and is gaining increasing attention of researchers from a broad disciplines such as engineering \cite{wang2021conditionsensenet}, biology \cite{ching2018opportunities}, and chemistry \cite{mater2019deep}. It can be expected that deep learning will be applied in more and more safety-critical applications like autonomous driving, avionics, and robotics \cite{fink2019deep} \cite{tzelepi2017human}, which typically require highly reliable processing to avoid catastrophic consequences. There have been intensive efforts devoted to enhance the robustness of deep learning against various perturbations like adversarial noise and natural noise in deep learning community to ensure safety deployment of deep learning \cite{biondi2019safe} \cite{tang2021robustart} \cite{rabe2021development}. In contrast, the influence of hardware faults in silicon-based computing fabrics such as deep learning accelerators, FPGAs, and GPUs that sustain efficient deep learning processing are generally overlooked. 

Deep learning models especially neural networks are known to be fault-tolerant inherently mainly because of the widely utilized  activation functions, pooling layers, and the ranking-based outputs that are usually insensitive to computing variations. Many prior work explored the inherent fault tolerance of neural networks for the sake of higher energy efficiency, performance, and memory footprint with approaches like voltage scaling \cite{xue2022winograd} \cite{paul2021voltage}, DRAM refresh scaling \cite{tu2018rana}, and low-bit-width quantization \cite{Zhao2020Linear} \cite{zhao2020bitpruner}. However, the unique fault-tolerant feature does not guarantee fault tolerance against hardware faults and even results in substantial accuracy variation across the different fault configurations according to the investigation in \cite{liu2021hyca} \cite{xu2020hybrid} \cite{zhang2019fault}, which essentially aggravates the uncertainty of the deep learning processing and hinders the deployment of deep learning in safety-critical applications. 

Prior approaches that are proposed to enhance the robustness of neural networks (NN) against the adversarial noise or natural noise can also guide fault-tolerant neural network processing against hardware faults in certain extent \cite{serban2020adversarial} \cite{tang2021robustart}, but the effectiveness can be limited because of the distinct mechanisms of influence on neural network processing. The perturbations caused by adversarial noise and natural noise only affect inputs of neural network processing while weights and neurons are generally intact. In contrast, hardware faults mostly pose more varied influence on neural network processing \cite{liu2021hyca} \cite{xu2020hybrid} \cite{zhang2018analyzing}. Specifically, faults in the on-chip buffers of the computing fabrics can affect not only neural network inputs but also weights and intermediate features of neural networks. Moreover, hardware faults can also be located at computing logic of neural network processing engines and corrupt the computing of neural network processing directly. As a result, it is usually more difficult to characterize the influence of hardware faults on neural network processing and they may pose distinct influence on neural network processing \cite{xu2020persistent} \cite{xu2021reliability} \cite{he2020fidelity} \cite{salami2018resilience}. Thereby, systematic fault-tolerant design approaches against various hardware faults remain highly demanded for the deployment of neural networks in safety-critical applications.

In order to mitigate the influence of hardware faults on neural network processing, a number of approaches from various angles have been proposed. Relevant surveys \cite{torres2017fault} \cite{shafique2020robust} \cite{khoshavi2020survey} \cite{mittal2020survey} mainly focus on fault-tolerant training approaches and many recent fault-tolerant approaches based on architectural design and circuit design are not classified or included. In this work, we investigate the fault-tolerant deep learning approaches from a hierarchical perspective that matches the general deep learning processing stacks all the way from high-level models to low-level circuits. With a top-down perspective, we have existing fault-tolerant deep learning design approaches divided into model layer, architecture layer, and circuit layer respectively. In addition, we observe that many approaches may cover multiple layers at the same time and briefly introduce the cross-layer approaches as well. Model layer fault-tolerant design approaches typically explore the inherent fault tolerance and redundancy in neural network models and have the models desensitized to computing variations and input variations induced by various hardware faults. Architecture layer fault-tolerant approaches mitigate hardware faults with specialized neural network accelerator architectures such as online recomputing \cite{liu2021hyca} and runtime voting \cite{gao2022soft}. Circuit layer fault-tolerant approaches focus on low-level circuit designs such as error-tolerant encoding, fine-grained modular redundancy \cite{mahdiani2012relaxed}, and stochastic circuits \cite{8493550}. Cross-layer fault-tolerant approaches usually combine fault-tolerant approaches from different layers in a unified framework to make best use of the different approaches and achieve more effective protection \cite{zhang2019fault}. 

The organization of this paper can be summarized as follows. Section \ref{sec:model} mainly introduces fault-tolerant model design approaches against hardware faults. Section \ref{sec:arch} focuses on fault-tolerant architectural design approaches particularly for deep learning accelerators. Section \ref{sec:circuit} mainly covers the circuit-based fault-tolerant design approaches with an emphasis on approximate computing and stochastic computing. Section \ref{sec:cross} briefly introduces the cross-layer design approaches. Section \ref{sec:conclusion} concludes this paper.
 %Introduction of this paper
\section{Model-layer Fault Tolerance} \label{sec:model}
\subsection{Related Work}
This section introduces the model-level fault tolerance techniques for NN applications, which lie at the top of the hierarchy. One can presume that low-level techniques (i.e., the architecture-level and circuit-level techniques) can provide better reliability guarantees since they could directly handle more types of actual hardware faults. 
Nevertheless, purely relying on low-level fault tolerance techniques for fault-tolerant deep learning can be prohibitively costly, especially considering the ongoing trend of scaling up the NN model capacities. Therefore, there exists a vast literature on exploiting the application-level or model-level characteristics of NNs to facilitate more economical fault-tolerant DL.

The key characteristic utilized by all model-level techniques is the NNs' inherent redundancy and tolerance for faults. In other words,
most NNs have only a fraction of neurons (sensitivity neurons), whose computational faults induce severe functional errors. Thus, all model-level techniques could be seen as designed revolving around neuron sensitivity. And we conclude them into three types, sensitivity analysis methods to assess neuron sensitivity, training strategies to alleviate or compensate for neuron sensitivities, and model architecture designs to eliminate or decrease the number of sensitive neurons.

\begin{figure*}[htb]
\begin{center}
\includegraphics[width=0.9\linewidth]{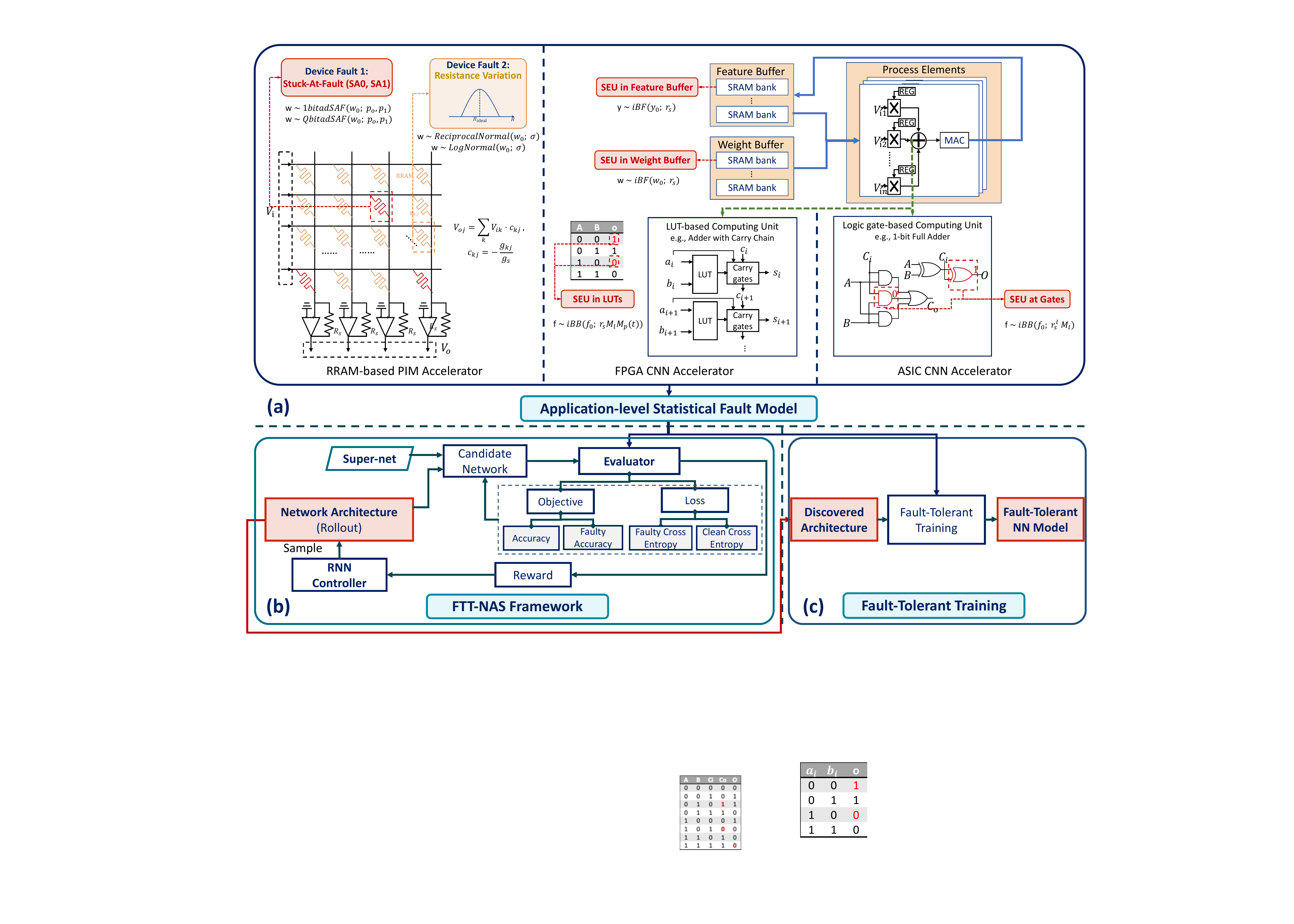}
\caption{The FTT-NAS workflow: (a) Establish algorithm-level fault models. (b) Parameter-sharing neural architecture search. (c) Fault-tolerant training of discovered architectures.}
\label{fig:ftt_workflow}
\end{center}
\end{figure*}

\subsubsection{Sensitivity Analysis}
To exploit the neural network (NN) applications’ characteristics for a more economical fault-tolerant DL solution, one should first understand the behavior of NN models with computational faults by conducting sensitivity analysis. Vialatte and Leduc-Primeau~\cite{vialatte2017astudy} analyze the layer-wise sensitivity of NN models under two fault models.
F. Libano et al.~\cite{libano2018selective} conduct layer-wise sensitivity analysis, and then propose to only triplicate the vulnerable layers, and thus reduce the triple modular redundancy (TMR) overhead for protecting an NN model. Christoph Schorn et al.~\cite{schorn2018accurate} propose a bit-flip resilience evaluation metric, and conduct sensitivity analysis of each individual neuron. The authors further refine their analysis model in \cite{schorn2019efficient}.
Guanpeng Li et al.~\cite{li2017understanding} find that the impacts and propagation of computational faults in an NN computation system depend on the hardware data path, the model topology, and the type of layers. 
These methods analyze the sensitivity of existing NN models at different granularities, and many of them also propose to exploit the analysis results to reduce the hardware overhead for reliability.

% On top of the previously mentioned bit-flip resilience evaluation metric~\cite{schorn2018accurate}, 

\subsubsection{Training Strategy}
% statistical fault-tolerant training
Fault-tolerant training is one of the commonest techniques to enhance the fault tolerance capability of an NN model.
In order to alleviate the influences induced by faults, many prior studies \cite{leung2012rbf} \cite{ahmed2011injecting} \cite{piotrowski2012comparison} \cite{cho2011node} \cite{osoba2013noise} \cite{hacene2019training} \cite{he2019noise} establish random weights or feature fault models, and inject faults accordingly during training. In this way, the NN models can learn to tolerate these types of faults. In other words, the sensitivities of neurons are alleviated such that their computational faults no longer lead to functional errors.

% analytical model based
Instead of statistically injecting faults into training, 
Christoph Schorn et al.~\cite{schorn2019efficient} propose a training strategy using analytical sensitivity analysis to adjust neuron sensitivities. This work argues that achieving a homogeneous resilience distribution inside the DNN can help obviate the need for special protection of critical parts. Accordingly, this work proposes an explicit weight rescaling technique to equalize the sensitivity metrics of different channels in one layer, and conducts the rescaling and finetuning processes iteratively.

% other regularizations
Other regularization techniques have also been proposed. As neural network outputs are usually more sensitive to large magnitude weights~\cite{bernier2000quantitative}, weight decay that limits the magnitude of weights is proposed to improve the fault tolerance \cite{huang2009advances}\cite{cavalieri1999novel}. Different from the training approaches with direct fault injection, Chi-Sing Leung et al.~\cite{leung2016regularizer} proposed a new objective function with an additional regularization term to minimize the training set errors and obtain fault-tolerant radial basis function (RBF) networks. It targets both weight fault and multiplicative weight noise. 

% retraining compensate for defunct neurons
Another type of studies~\cite{liu2017rescuing}\cite{xia2017fault}\cite{xu2019resilient}\cite{zhang2019fault} design retraining strategies to compensate for the performance loss caused by defunct neurons. These methods retrain the NN model to restore the model performance, after the online detection of the actual faults and variations. %J. Zhang et al.~\cite{zhang2019fault} propose to bypass faulty PEs by pruning and then recover the model performance with retraining. 

\subsubsection{Model Architecture Design}

As the output neurons directly influence the output, Tao Liu et al.~\cite{liu2019fault} propose to use error-correcting output codes (ECOC)~\cite{dietterich1994solving} to tolerate variations and SAFs. Specifically, they replace the conventional softmax with a collaborative logistic classifier that leverages asymmetric binary classification coupled with an optimized variable-length decode-free ECOC.

Ching-Tai Chiu et al.~\cite{Training1994Ching-Tai} propose to add additional hidden nodes to avoid model accuracy loss and repeatedly remove nodes that do not significantly affect the network output. A. Ahmadi et al.~\cite{Ahmadi2009low-cost} proposed to add a spare neuron which can be reconfigured to compare with any neuron in the model. It can be used for both fault detection and correction, but it is limited to spatial neural network architecture and can only be used to recover from single faults.
FTSET~\cite{ftset} uses simulation to analyze the sensitivity and replicates the critical neurons. Christoph Schorn et al.~\cite{schorn2019efficient} conduct some simple manual architecture modifications to eliminate sensitive neurons identified by analytical analysis.

Apart from these manual architecture designs and modifications, a recent work~\cite{fttnas} proposes FTT-NAS, which adopts the neural architecture search (NAS) technique to automate the process of finding a more fault-tolerant model architecture. We give a case study on FTT-NAS as follows.

\subsection{Case Study: Automated Architecture Search for Fault Tolerance}
\subsubsection{The FTT-NAS Workflow}
\begin{figure}[ht]
\centering
\subfloat[]{\includegraphics[width=0.64\linewidth]{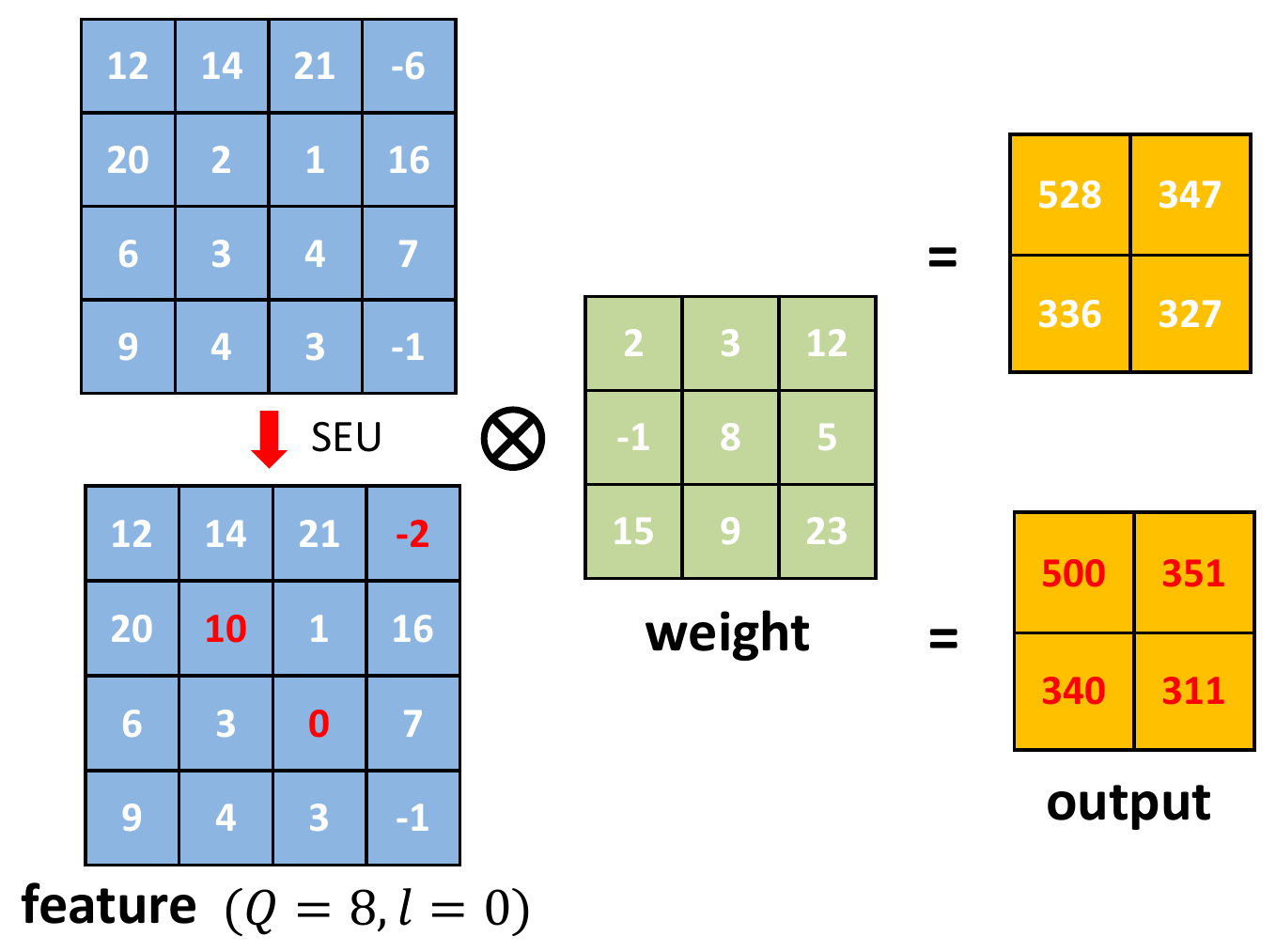}
}
\hfil
\subfloat[]{\includegraphics[width=0.88\linewidth]{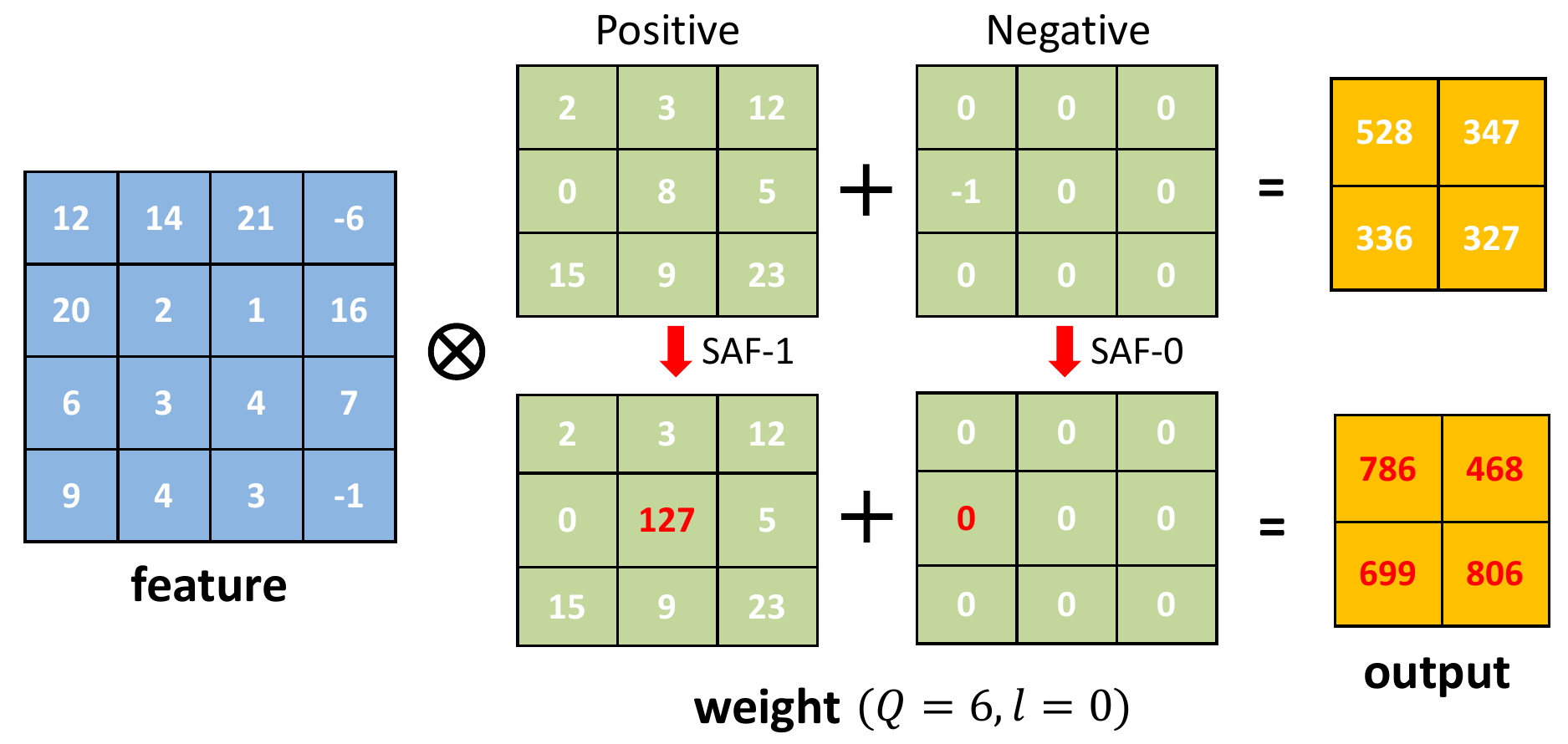}
}
\caption{Example of injecting faults. (a) Feature faults under the iBB fault model (soft errors in FPGA LUTs). Random bit-flip in the adder tree LUTs caused a power-2 bias to be added or subtracted from the feature value. (b) Weight faults under the adSAF fault model (SAF errors in RRAM cells). SAF in the memristor cells caused the weight value to be stuck at the boundary of the quantization representation range. Note that the usual NN to RRAM crossbar mapping scheme uses two crossbars to separately store positive and negative weights.}
\label{fig:ftt_fault_example}
\end{figure}

FTT-NAS aims at improving the NN model's algorithmic fault tolerance from the architectural perspective. The overall workflow is shown in Fig.~\ref{fig:ftt_workflow}. Firstly, in order to evaluate the fault tolerance of an NN model efficiently, FTT-NAS abstracts the algorithm-level fault model for injection-based evaluation.
Specifically, the authors analyze the convolution computations on different types of NN accelerators, and conclude two representative weight and feature fault models: the MAC-i.i.d Bit-Bias (MiBB) feature fault model, and the arbitrary-distributed Stuck-at-Fault (adSAF) weight fault model. The MiBB feature fault model abstracts the faulty effects of the feature map caused by random bit-flips in FPGA LUTs constructing the adder tree. And the adSAF weight fault model corresponds to the stuck-at faults occurring in the memristor cells of the RRAM crossbar. We show the examples of injecting these two types of faults into the convolution computation in Fig.~\ref{fig:ftt_fault_example}.

Secondly, according to preliminary experiments on what types of architectural decisions influence the fault tolerance capability, FTT-NAS designs a large search space containing about $10^{25}$ architectures. %This architectural space has a similar macro layout and design space of connection topologies with existing cell-based search spaces~\cite{enas,darts}, and has up to 11 types of primitives to be chosen for each connection. 
Then, FTT-NAS employs parameter-sharing NAS~\cite{enas} to search for fault-tolerant architecture in this search space. Specifically, FTT-NAS constructs an over-parameterized super network that contains the parameters needed to evaluate all architectures in the search space, and an RNN-based controller to sample architecture from the search space. The weights of controller are updated using the reward evaluated using the super network. And FTT-NAS inject faults according to the previously established fault model into the training and evaluation process of the super network.
Finally, FTT-NAS derives a final architecture using the controller, and trains it with fault-tolerant training.

\subsubsection{Sample Results} 
\begin{figure*}[ht]
\begin{center}
\includegraphics[width=0.95\linewidth]{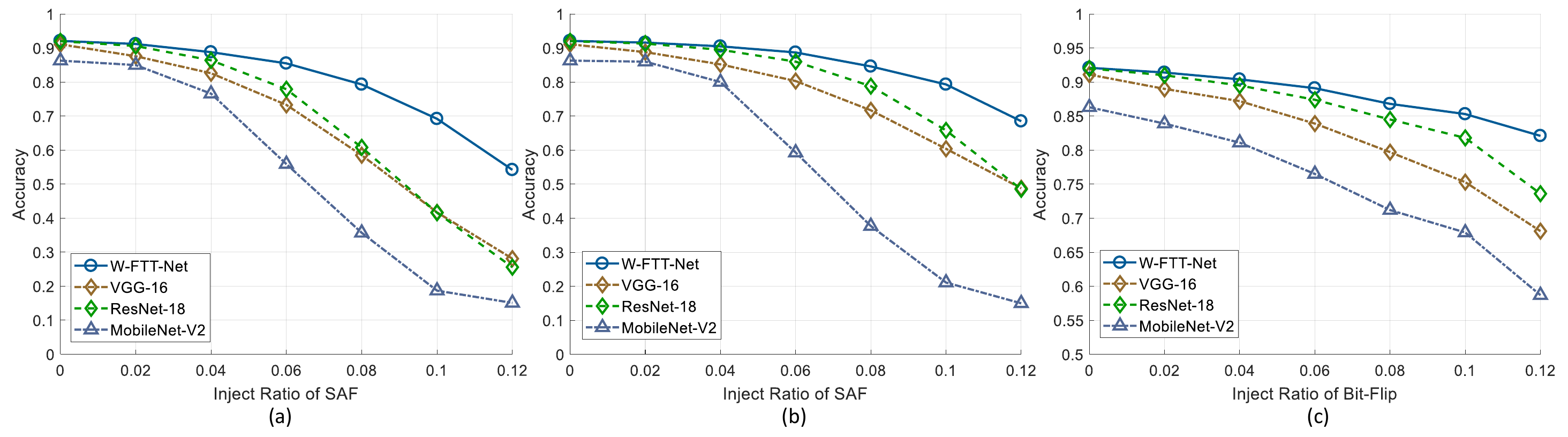}
\caption{Accuracy curves under different weight fault models. (a) 8bit-adSAF model (caused by SAFs in 8-bit RRAM memristor cells). (b) 1bit-adSAF model (caused by SAFs in 1-bit RRAM memristor cells). (c) iBF model (caused by random bit flips in the weight buffer).}
\label{fig:ftt_weight_results}
\end{center}
\end{figure*}

FTT-NAS compares the fault tolerance of its discovered architectures with baseline architectures, including ResNet, VGG, MobileNet. 
For example, Fig.~\ref{fig:ftt_weight_results} shows the reliability comparison between the baseline architectures and the W-FTT-Net architecture discovered under the 8bit-adSAF fault model. During the test time, the reliability is evaluated using three different types of weight fault models, including the 8bit-adSAF model, the 1bit-adSAF model, and the iBF model. As we can see, W-FTT-Net outperforms baseline architectures consistently at different noise levels under three different types of weight fault models. 

%\subsubsection{Results}

% Model design
% Model retraining
% Activation constraints
% Stochastic neural networks
% Combined architectures %fault tolerant model design
\section{Architecture-Layer Fault Tolerance} \label{sec:arch}
Despite the inherent fault tolerance of neural networks, it is insufficient to protect the neural network processing with only model layer design because of the vast fault configuration space that can hardly be fully considered during the model design stage. Moreover, model layer fault-tolerant design approaches typically require time-consuming training, which inhibits the runtime fault recovery. Moreover, training usually relies on application data which may not be always accessible during the deployment stage. Thereby, fault-tolerant architectural designs which can potentially mitigate hardware faults with much less limitation on neural network models and high-level applications are investigated recently. 

\subsection{Related Work}
While convolution in neural networks can be viewed as matrix-matrix multiplication, some of the fault-tolerant architectural designs are inspired by fault-tolerant matrix-matrix multiplication methods such as algorithm-based fault tolerance (ABFT) proposed in \cite{huang1984algorithm}. The basic idea is to add sum of each row/column of the input matrices to the original input matrices such that the sum of output matrices in each row/column can be obtained from both the extended matrix-matrix multiplication and accumulation of elements in the original output matrix. In this case, checksum of the results from different approaches can be performed to detect errors in the matrix-matrix multiplication. At the same time, single bit error can also be recovered based on the checksum on both row and column accumulation. E. Ozen et al. \cite{ozen2019sanity} took advantage of the regular computing pattern of convolution neural networks and applied the ABFT technique to protect neural network accelerators against soft errors. Kai Zhao et al. \cite{zhao2020ft} explored the various data flows of ABFT technique for convolution operations to obtain the fault detection/correction capability comprehensively, which can also be utilized to guide fault-tolerant neural network accelerator designs. While the overhead of the naive ABFT is non-trivial, Dionysios Filippas et al. \cite{filippas2021low} proposed a lightweight ABFT implementation, ConvGuard, which predicts the output checksum of convolution implicitly by accumulating only the pixels at the border of the dropped input features. Thibaut Marty et al. \cite{marty2020safe} proposed to utilize the ABFT technique to mitigate timing errors induced by overclocking of the neural network accelerators on FPGAs. Their experiments reveal that the proposed ABFT design poses negligible area overhead, enables aggressive overclocking of the neural network accelerators, and achieves up to 60\% throughput improvement of the overall neural network processing. 

Unlike the ABFT-based architectural designs that are generally independent with the specific deployed neural networks, many approaches also explore the features of neural networks and develop corresponding fault-tolerant architectural design to achieve more effective protection. Christoph Schorn et al. \cite{schorn2018accurate} investigated the importance variations of neurons in the models and proposed an heterogeneous computing array that provides two different levels of fault tolerance. In this case, important neurons are allocated to highly resilient computing array partitions while less important neurons are allocated to the rest of computing array, which ensures resilient neural network processing with much less hardware overhead. Jeff Zhang et al. \cite{zhang2018analyzing} \cite{zhang2019fault} proposed to add a zero bypass data path to processing elements (PEs) in neural network accelerators and the bypass will be enabled when the corresponding PEs are faulty. Although zero bypassing typically has less yet predicted influence on the neural network processing compared to values with random faults, it may still cause substantial accuracy loss. To address the problem, the authors perform retraining to adapt to each specific fault configuration, which essentially alters the importance of neural network weights or neurons to suit the faulty computing array. Navid Khoshavi et al. \cite{khoshavi2020shieldenn} proposed an online fault assessment paradigm to delineate the most vulnerable parts of neural networks. On top of the assessment, they provided corresponding hardening strategy to accomplish optimized neural network accelerator designs against transient errors with resource constraints.

In addition, there are also conventional fault-tolerant computing mechanisms closely combined with neural network accelerator architectures. Zhen Gao et al. \cite{gao2022soft} introduced ensemble learning to fault-tolerant neural network processing for the first time and combined it with redundancy design. The basic idea is to have a group of redundant base neural network models implemented in parallel and equipped the different implementations with a score comparison voter on FPGAs such that the majority of the soft error induced prediction errors can be mitigated with negligible hardware overhead. The different base models are all lightweight compared to the large original model, so the overhead is much smaller compared to conventional triple modular design (TMR) on the original model. Meanwhile, the ensemble model that is combined on top of the lightweight models can also achieve competitive accuracy compared to large-scale models. Cheng Liu et al. \cite{liu2021hyca} \cite{xu2020hybrid} proposed to apply conventional recomputing mechanism to fault-tolerant neural network accelerator designs with a hybrid computing architecture. The basic idea is to have additional computing units seated along with a classical neural network accelerator to recompute the operations mapped to the faulty PEs. Since the additional computing units have each faulty operations processed in parallel independently, the recomputing fabric can be utilized to fix faulty neural network accelerators with arbitrary distribution of the faulty PEs. 

In summary, there have been a number of fault-tolerant neural network accelerator designs proposed from distinct architectural angles. They differ in terms of transparency to neural network models, target hardware fault types, hardware overhead, and performance penalty. There is no determined answer for all the fault-tolerant requirements and the brief survey can be utilized to guide the optimized fault-tolerant architecture selection. 

\subsection{Case Study: Hybrid Computing Architecture for Deep Learning Accelerators}
In this sub section, we will take the hybrid computing architecture (HyCA) proposed in \cite{liu2021hyca} as a case study and illustrate how it can be utilized as a general architecture for fault-tolerant neural network processing.

\subsubsection{HyCA Architecture}
\begin{figure}
	\center{\includegraphics[width=0.85\linewidth]{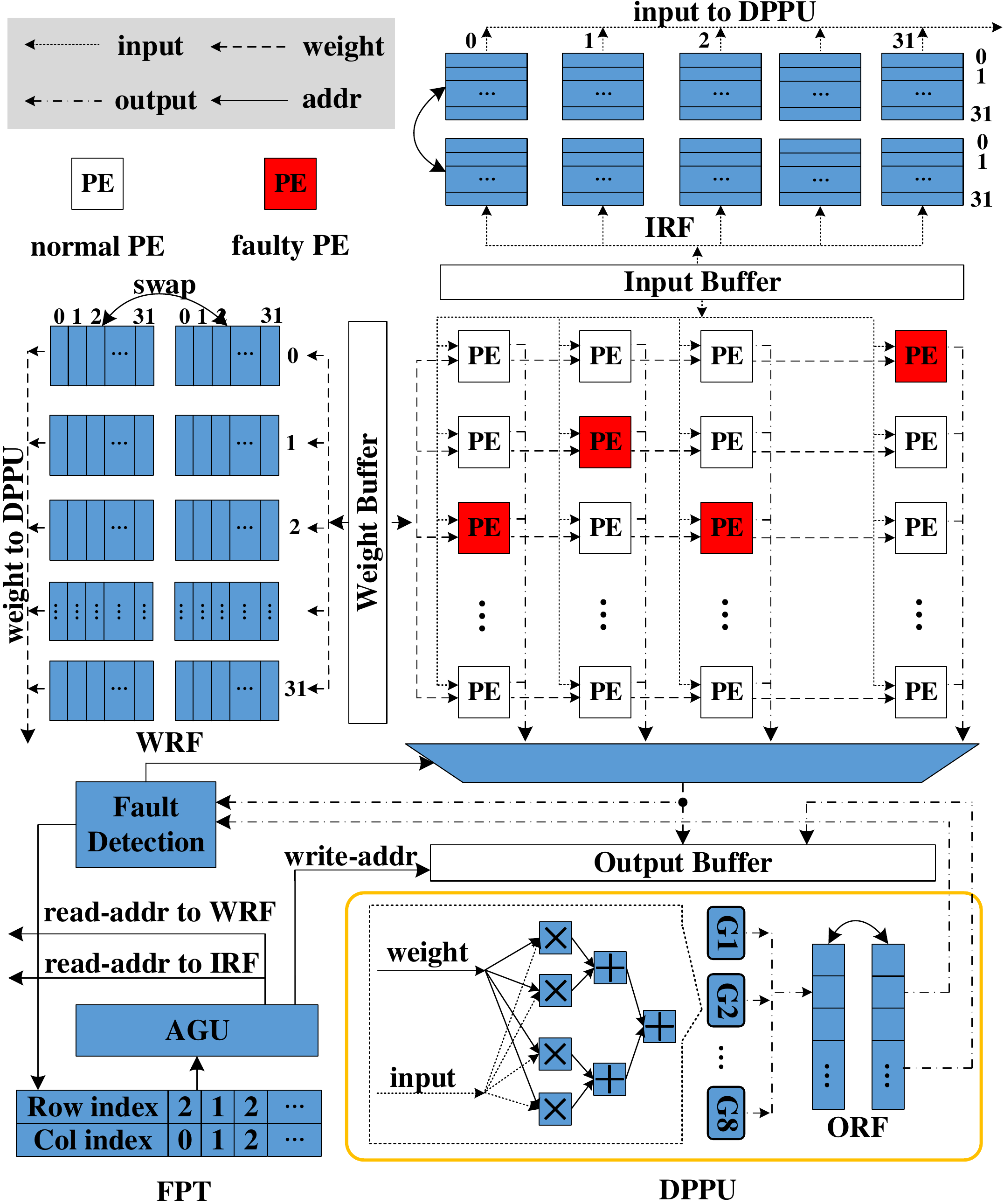}}
    \caption{Overview of of a DLA with Hybrid Computing Architecture. The components highlighted with blue are added to the conventional DLA to tolerate faulty PEs in arbitrary locations of the 2-D computing array.}
\label{fig:mapping}
\end{figure}

Figure. \ref{fig:mapping} presents an overview of HyCA for fault-tolerant neural network processing. It has a dot-product processing unit (DPPU) seated along with a classical 2-D computing computing array, to recompute all the operations mapped to the faulty PEs in arbitrary locations of the 2-D computing array. While the 2-D computing array has each PE to calculate the different output features sequentially given the output stationary data flow \cite{Chen2016Eyeriss} and the DPPU has all the PEs to compute a single output features in parallel. 

To make sure that the normal 2-D array processing will not be affected by the DPPU recomputing, DPPU cannot read the required weights and input features aligned in channel dimension if it starts the recomputing at the same time with the 2-D computing array. To that end, we have the input features and weights buffered in an input register file (IRF) and a weight register file (WRF) respectively while they are read for the 2-D computing array processing. Meanwhile, we have the recomputing delayed until there are sufficient inputs and weights ready for the recomputing. Accordingly, the delay must be larger than or equal to the number of weights required by DPPU data consumption in a single cycle to ensure DPPU can be fully utilized. As the DPPU may recompute operations on any PE in the 2-D computing array, the delay also needs to be larger than or equal to $Col$ when the last column of the PEs obtain the weights passed from the first column of PEs. Note that $Col$ refers to the column size of the 2-D computing array. 

In this work, we organize IRF and WRF in Ping-Pong manner to ensure that the 2-D computing array can continue the normal dataflow without any stall during the DPPU recomputing. As the DPPU conducts the output feature calculation in parallel, DPPU can always finish the recomputing of the operations mapped to the faulty PEs before the Ping-Pong register files swap with each other when the DPPU size does not exceed the number of the faulty PEs. Note that DPPU size refers to the number of multipliers in DPPU. Since the peak computing power of DPPU equals to that of the 2-D computing array when configured with the same number of PEs, DPPU size is comparable to the 2-D computing array size and can also be used to represent its computing power. This also explains why DPPU can always finish the recomputing tasks before new weights and inputs are ready when DPPU size is larger than the number of faulty PEs in the 2-D computing array. 

In addition, we have a fault PE table (FPT) to record the coordinates of the faulty PEs in the 2-D computing array which can be usually obtained with a power-on self-test procedure. With the coordinates of faulty PEs, an address generation unit (AGU) is used to generate the read addresses and instruct the DPPU to read the right input features and weights from the register files. Moreover, AGU also determines the addresses to the output buffer for the overlapped writes of the recomputed output features. Similar to the IRF and WRF, there is also a Ping-Pong register file called ORF for the DPPU outputs and it is utilized to pipeline the DPPU recomputing and the write from DPPU to the output buffer.

\subsubsection{Experiments}
In this experiment, we have two different fault distribution models including the random distribution model and the clustered distribution model implemented. For the random distribution model, the faults are randomly distributed across the entire computing array. For the clustered distribution model which is usually used to characterize the manufacture defects, the faults are more likely to be close to each other and the model proposed in \cite{meyer1989modeling} is applied in this work. Meanwhile, we notice that the influence of hardware faults is related with the fault distribution, so we generate 10000 configurations randomly for each fault injection rate and average the evaluation in the experiments. Since we mainly focus on the reliability of the regular 2-D computing array in a deep learning accelerator, we use PE error rate (PER) as the fault injection metric similar to the work in \cite{zhang2018analyzing} and \cite{qian2016optimal}. We evaluate the hard error rate in a large scale ranging from 0$\%$ to 6$\%$. 

\begin{figure}
	\center{\includegraphics[width=1\linewidth]{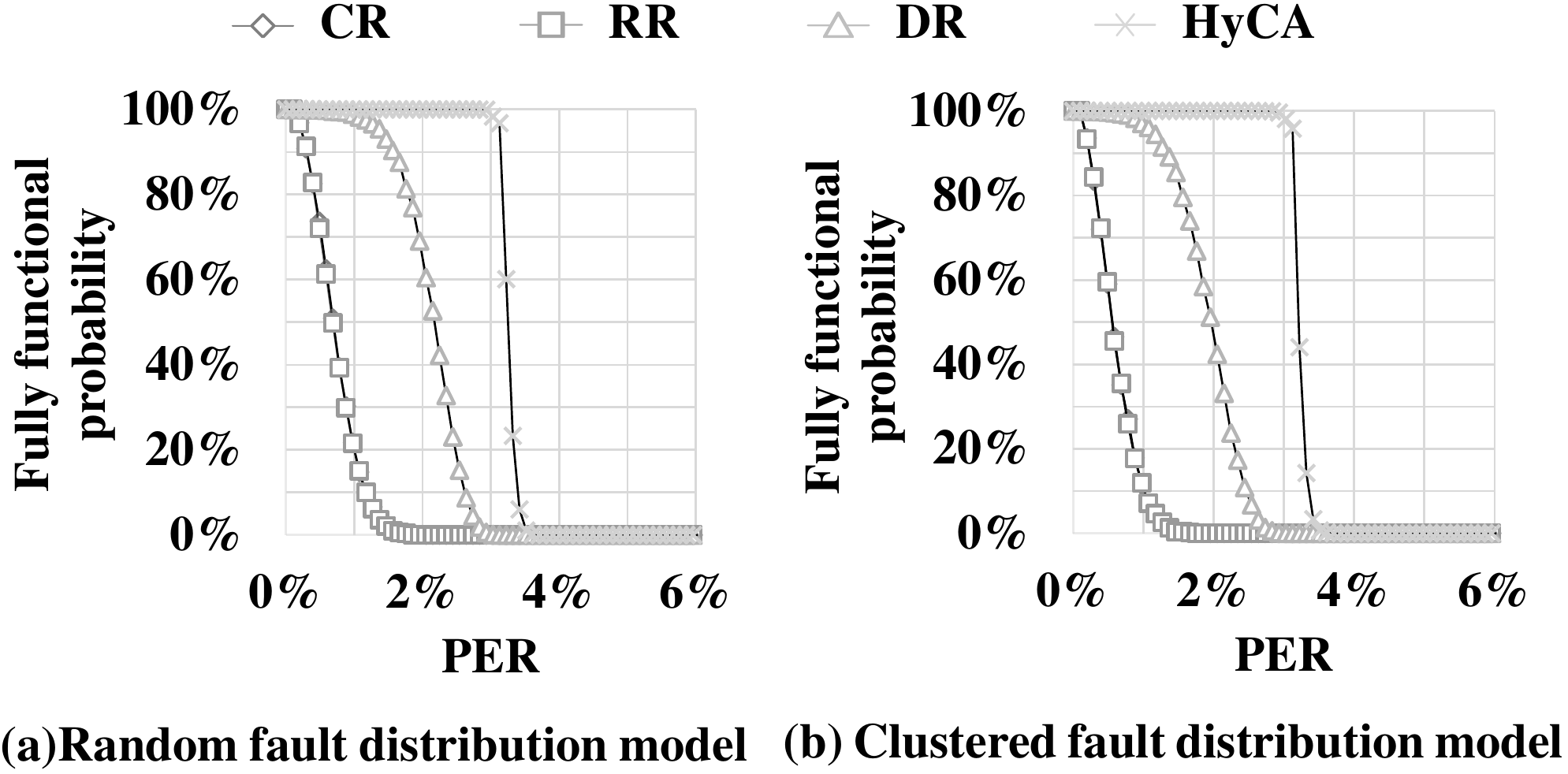}}
    \caption{Fully functional probability of DLAs with different redundancy approaches.}
\label{fig:survival}
\end{figure}

To evaluate the reliability of the DLAs, we propose two metrics that can be applied for different applications. One of them is the fully functional probability and it shows the probability that the DLA can be fully functional without any performance penalty. It is preferred by the mission-critical applications that do not allow any performance degradation nor model modification because any system modification may require expensive and lengthy safety evaluation and certification. The experiment is shown in Figure. \ref{fig:survival}. It shows that HyCA outperforms the three classical redundancy approaches and the advantage gets enlarged under the clustered fault distribution. The main reason is that each redundant PE in row redundancy (RR), column redundancy (CR) and diagonal redundancy (DR) can only be utilized to replace a single faulty PE in a row, a column, and a row-column pair respectively. When multiple faults occur in the same protected region, these redundancy approaches fail to recover the faulty 2-D computing array and the design will not be fully functional. Unlike these classical redundancy approaches, HyCA allows arbitrary faulty distribution and can perfectly repair the computing array as long as the number of faulty PEs in the 2-D computing array does not exceed the DPPU size. Thereby, the fully functional probability of HyCA is not sensitive to the fault distribution models. 

\begin{figure}
	\center{\includegraphics[width=0.99\linewidth]{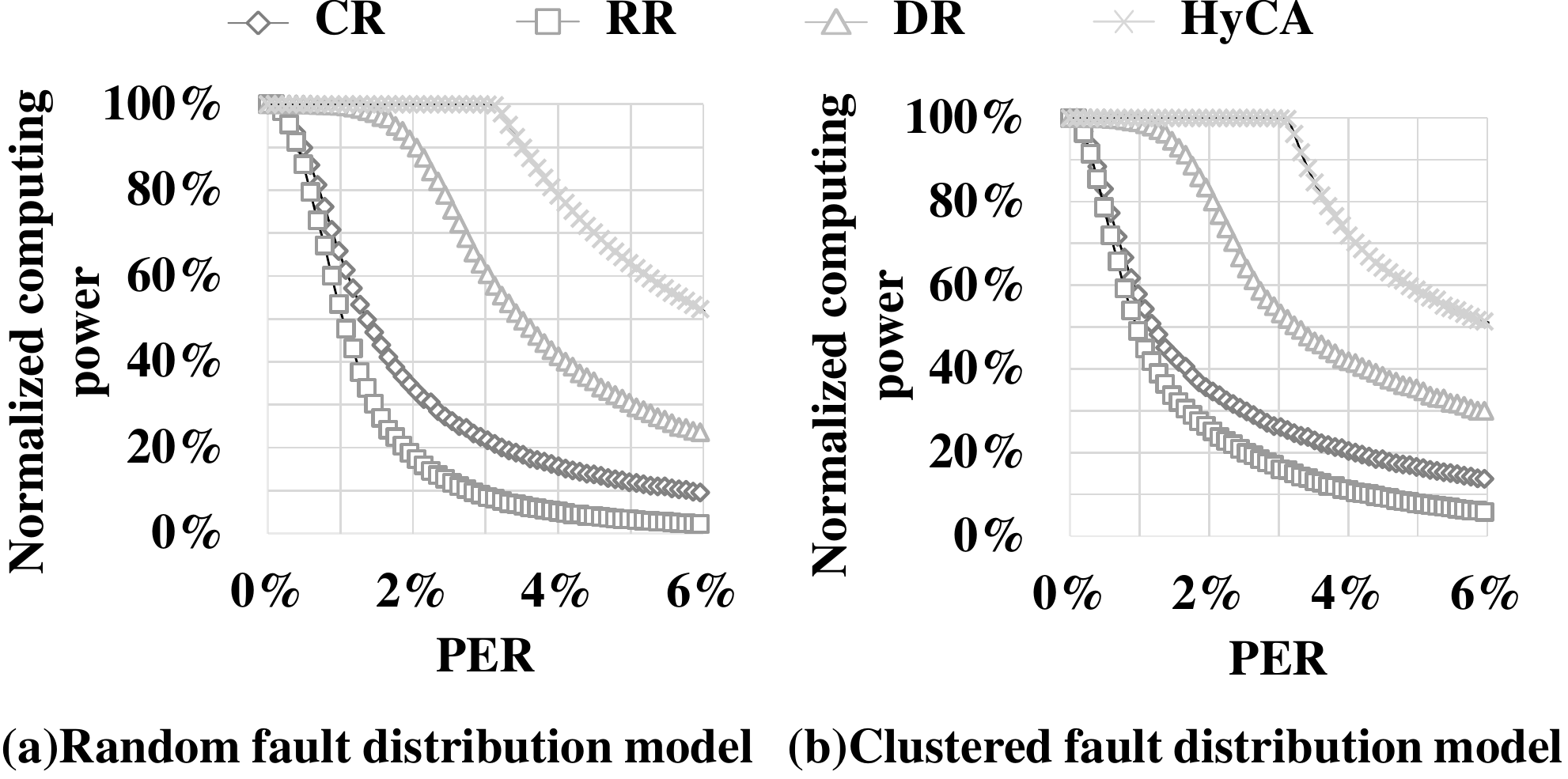}}
    \caption{Normalized computing power of DLAs with different redundancy approaches.}
\label{fig:available}
\end{figure}

The other metric is the normalized remaining computing power and it refers to the percentage of the remaining computing array size over the original 2-D computing array size. This metric is particularly important for the non-critical applications that do not require fully functional accelerators and allow the accelerators to be degraded, because the remaining computing array size determines the theoretical computing power and affects the performance of the deployed neural network models directly. Figur. \ref{fig:available} reveals the computing power comparison of the different redundancy approaches. It can be observed that HyCA shows significantly higher computing power under all the different PER setups and the advantage also enlarges with the increase of the PER. This is mainly brought by the fault recovery flexibility of the HyCA that allows the DPPU to select the most critical faulty PEs to repair when the redundant faulty PEs are insufficient. In contrast, each redundant PE can only repair a limited subset of the faulty PEs for the RR, CR and DR. There is little space left to optimize the faulty PE mitigation order. Thereby, the remaining computing power of RR, CR, and DR is much lower. 

\subsubsection{Summary}
Prior redundancy design approaches for the regular computing array such as RR and CR greatly reduce the hardware overhead compared to the classical TMR approaches, but they are rather sensitive to the fault distribution and fail to work especially when the faults are unevenly distributed. HyCA has a DPPU to recompute all the operations mapped to the faulty PEs in the 2-D computing array. When the number of faulty PEs in the 2-D computing array is less than the DPPU size, HyCA can fully recover the 2-D computing array despite the fault distribution. Even when the fault error rate further increases, DPPU can still be used to repair the most critical PEs first to ensure a large available computing array and minimize the performance penalty.  %fault tolerant architecture design
\section{Circuit-layer Fault Tolerance} \label{sec:circuit}
Hamid Reza Mahdiani et al. \cite{mahdiani2012relaxed} proposed to take advantage of the inherent fault tolerance in neural network applications by using relaxed fault-tolerant processing elements for neural network processing. Instead of conducting PE-level TMR protection, this approach enables selective ripple-carry adder cell TMR protection for the PEs in neural network accelerators based on the affected bit position of the outputs. Basically, cells that affect the higher output bits will be protected with higher priority and the exact protection strategy depends on the corresponding fault tolerance level requirements of the applications. The experiment reveals that this approach shows much less chip area and shorter critical path compared to fully TMR protection under the same protection level. However, compared to the computing datapath, memory access is more energy-consuming \cite{6757323}. In \cite{7551399}, a framework, ``Minerva'', is developed to optimize a neural network accelerator and the energy due to SRAM is reduced by scaling its supply voltage. The power consumption reduces quadratically with the supply voltage while the bit error rate increases exponentially. For a minimal accuracy reduction caused by the bit errors, Razor SRAMs and bit/word masking are employed to detect and correct the errors respectively. Consequently, a fault probability of more than $10^{-3}$ can be tolerated and 2.7$\times$ power saving is achieved. Similarly, Lita Yang et al. \cite{yang2018bit} explored the fault tolerance of binary neural networks and took advantage of the fault tolerance to reduce the SRAM voltage in a convolutional neural network (CNN) processor. The experiment reveals significant energy savings with limited accuracy degradation, though it may vary across the network topologies and classification tasks. Juan Antonio Clemente et al. \cite{clemente2016hardware} developed a fault-tolerant Hopfiled neural network (FT-HNN) on FPGAs by inserting additional connection to obtain hidden accumulation states differently and then had them voted for fault tolerance. Since the different accumulations share many partial results, the proposed FT-HNN requires much less hardware overhead compared to the baseline TMR implementation (HNN+TMR), but FT-HNN still shows comparable standard errors and convergence over HNN+TMR under the same SEU setup.  

Emerging computing paradigm, such as approximate computing and stochastic computing, can also exploit the fault-tolerant feature of many applications, so that higher performance and energy efficiency are obtained with limited accuracy loss.
\subsection{Approximate computing circuits}
Approximate computing has widely been investigated on various levels of a computing system, from the programmable language and algorithm down to the circuit. On circuit level, commonly used approximation techniques include the voltage overscaling (VOS)~\cite{vos2013}, implementing a complex arithmetic operation based on a simplification of its mathematical representation~\cite{pam2022}, and modification of the classical (accurate) logic function of an arithmetic circuit~\cite{divider2019}. As the most simple and convenient approach, VOS reduces the power dissipation of computing without the need of circuit modification. However, it may result in uncertain errors in the more significant bits of arithmetic operations, which can damage the accuracy of the entire system. BY modifying the classical design of an arithmetic circuit, the hardware overhead lowered with deterministic errors; thus, the errors can be characterized and/or compensated in the following computations. Considering deep learning applications, due to the recurrent refinement of learning algorithms, some errors of arithmetic circuits can be recovered at the system level. As a result, approximate arithmetic circuits have been potential choices for implementing a hardware-efficient deep learning algorithm. 

\cite{kws2020} proposes a precision self-adaptive approximate addition unit in the design of a binarized weight network processor for keyword-spotting (KWS). Compared to the accurate computing mode, the approximate addition reduces the power consumption by $1.3\times$. In addition, approximate multiplication is utilized in the required mel-scale frequency cepstral coefficients module. Finally, the power dissipation of the KWS system is reduced by $1.4\times$ with a less than 0.5\% recognition accuracy loss.

In \cite{9165786}, approximate arithmetic circuit designs including adders, multipliers and dividers based on various approximation methodologies are introduced. These designs are comprehensively evaluated by different metrics with respect to their accuracy and hardware efficiency. Moreover, explorations are performed to reveal the relationship between the statistical error metric of approximate arithmetic circuits and the accuracy of image processing and deep neural network applications using them. Specifically, by using various approximate adders and multipliers, several image processing algorithms and the face detection and alignment implemented by using a multi-task CNN are performed. The simulation results show that for simple operations without many serial computations, such as a sum of product in an image filtering, the approximate arithmetic circuits with smaller mean relative error distances (MREDs) generally leads to higher quality. The MRED is defined as 
\begin{equation}
\textrm{MRED}=\frac{\sum_{i=1}^N|\frac{\hat{x}_i-x_i}{x_i}}|{N}.
\end{equation}  
$\hat{x}_i$ represents the approximate result and $x_i$ the real value.

In deep learning, as complex computations such as multiple of consecutive matrix multiplications, except for MRED, error bias is of great importance. The simulation results show that approximate adders and multipliers with small error bias general result in low degradation in the accuracy of face detection and alignment. The error bias is given by
\begin{equation}
\textrm{Error bias}=\frac{\sum_{i=1}^N{(\hat{x}_i-x_i)}}{N}.
\end{equation} 
The simulation results also show that, with a same bit width, deep learning application is more sensitive to the errors of adders than those of multipliers. Another interesting conclusion from \cite{9165786} is that, by using approximate arithmetic circuits, deep learning applications can achieve benefits in both energy-efficiency and accuracy.

\subsection{Stochastic computing circuits}
\begin{figure}
	\centering
	\includegraphics[width=0.35\textwidth]{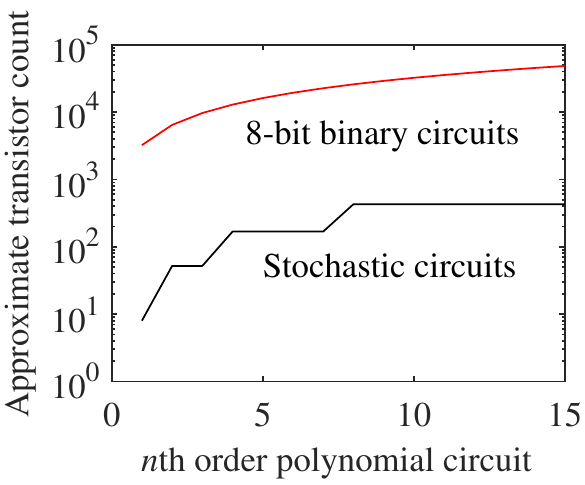}
	\caption{Transistor count implementing Bernstein polynomials with different order, stochastic computing vs. 8-bit binary circuits.}
	\label{fig:trans_count}
\end{figure}

Stochastic computing (SC) is an alternative computing paradigm that produces unbiased estimate of the actual results, i.e., the error bias is 0. In SC, numbers are encoded by streams of random 0s and 1s and the probability of 1 in this sequence is used to represent a number. For example, ``10011100'' can be used to represent 0.5 in the unipolar representation, where the encoded number equals to the probability of 1 in the sequence. Using SC, the complexity of the arithmetic circuits can be greatly reduced. For example, an AND gate implements a multiplier in the unipolar representation in SC. The output of an AND gate is 1 only when both the inputs are 1s. So the probability of the AND gate generating a 1 is the product of the probabilities of 1s of the two input sequences given that they are independently generated. Nevertheless, in conventional binary circuits, it typically takes hundreds of gates to build a multiplier. Fig. \ref{fig:trans_count} shows the transistor count of core circuits computing Bernstein polynomials using SC compared to the conventional 8-bit binary counterpart. The SC circuits are synthesized by the method proposed in \cite{5601694}. The results show that stochastic computing circuit can achieve one hundredth of the hardware cost of its 8-bit binary counterpart.  

\begin{figure}
	\centering
	\includegraphics[width=0.45\textwidth]{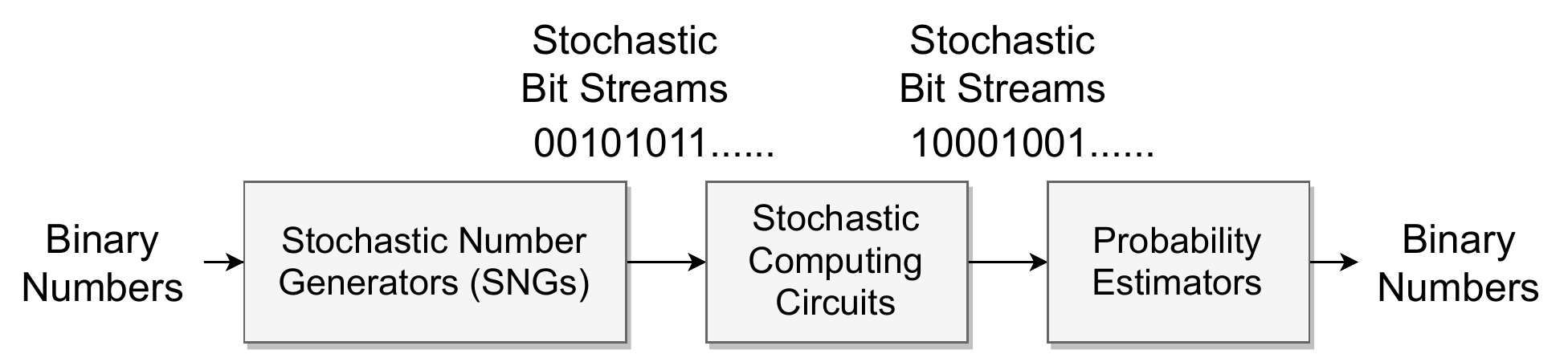}
	\caption{A typical stochastic computing system contains stochastic number generators (SNGs), stochastic computing circuits and probability estimators.}
	\label{fig:sto_sys}
\end{figure}

However, the peripheral supporting circuits would take a large portion of an SC system since the stochastic bit stream is typically generated by the costly stochastic number generators (SNGs) and later converted back to binary numbers by probability estimators, as shown in \ref{fig:sto_sys}. In the unipolar representation, the stochastic bit stream can be generated by comparing the number to be encoded, $x$, with a uniformly distributed random number. If $x$ is larger, a `1' is generated; otherwise, a `0' is produced. So the probability generating a `1' equals to $x$. The final output can be estimated by counting the number of 1's in the output bit stream and divide it by the sequence length.

In \cite{8493550}, the SC accelerators are studied from an architectural perspective and the conversion cost can be mitigated by sharing the components of the conversion units across the SC circuits. It is also found that a higher computation-to-conversion ratio indicates a lower conversion overhead and thus a higher energy efficiency gain. This is often the case in compute-intensive tasks such as convolution, which is the major operations in a CNN. For a 3$times$3 Gaussian filter and 5$times$5 general convolution, the conversion overhead can be shared by 6.6 to 18.2 arithmetic operations by exploiting data locality and reuse. Eventually, the SC circuits achieve higher energy efficiency than their binary counterparts at a bit precision of 8 or lower with limited accuracy degradation of within 1\% for an SVM classification task. 

\begin{figure}
	\centering
	\includegraphics[width=0.45\textwidth]{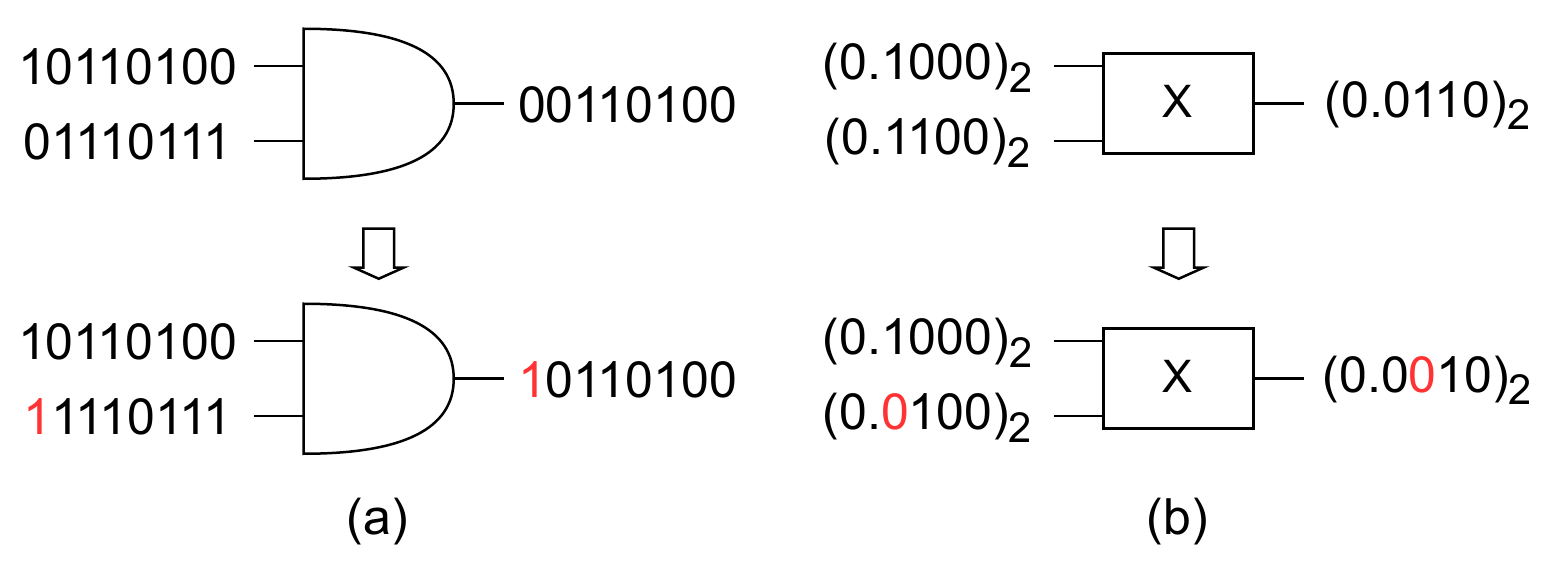}
	\caption{A typical stochastic computing system contains stochastic number generators (SNGs), stochastic computing circuits and probability estimators.}
	\label{fig:SC_FT}
\end{figure}

When considering the inherit fault-tolerant nature of SC, larger energy savings can be achieved by voltage scaling. This stems from the unique coding method of SC, i.e., numbers are encoded as equal-weighted long bit streams and each bit only accounts for $1/L$ of its value when the total sequence length is $L$. Therefore, one or few bit flip error in the bit stream does not affect the final results very much. On the contrary, in the binary systems, the numbers mostly use positional coding, i.e., significant bits usually have larger weights. Thus, a bit flip error on the most significant bit can lead to a large error, as compared in Fig. \ref{fig:SC_FT}.

Due to this fault-tolerance feature, voltage scaling can be applied to SC circuits to further reduce the energy cost. However, when the supply voltage is beyond its critical point, timing violations and bit flips may occur. \cite{8493550} shows that on a fabricated ASIC prototype, the SC circuits can tolerate these errors and operate at a supply voltage as low as 0.55 V while producing satisfying results, whereas the binary circuits fail at 0.8 V under the same working condition. This can bring an extra 3.3$times$ energy improvement for the SC circuits. 

In the context of conventional binary system, binarized neural networks (BNNs) can also exploit bit-wise computations to perform multiplications and additions using XNOR gates and bit-counting respectively \cite{rastegari2016xnor}. However, the real values are deterministically binarized through a comparison with 0. As a result, a bit flip can change the results significantly. \cite{9605019} compares the fault-tolerance under different bit-flip rates of SC-based neural networks (SCNNs) and BNNs. The results show that at different bit-flip rate levels, the recognition accuracy of SCNNs is always higher than that of the BNNs for the MNIST dataset. The accuracy of SCNNs falls below 97\% when the bit-flip rate is higher than 10\% while the number is at most 5\% for the BNNs when both the activations and weights are affected by the bit-flip errors. Conventional fault-tolerant schemes such as modular redundancy can be applied to BNNs to reduce its vulnerability to noise. However, the recognition accuracy of BNNs still cannot match that of an SCNN for the MNIST dataset. But when tested on a CNN for the CIFAR10 dataset, BNNs with modular redundancy outperform the SCNN, but at the cost of extra hardware resources. On the other hand, the long latency is a drawback of SCNNs due to the requirement for a long stochastic bit stream to perform computations with acceptable accuracy. It also increase the energy consumption of the SC circuits.  

\begin{figure}
	\centering
	\includegraphics[width=0.45\textwidth]{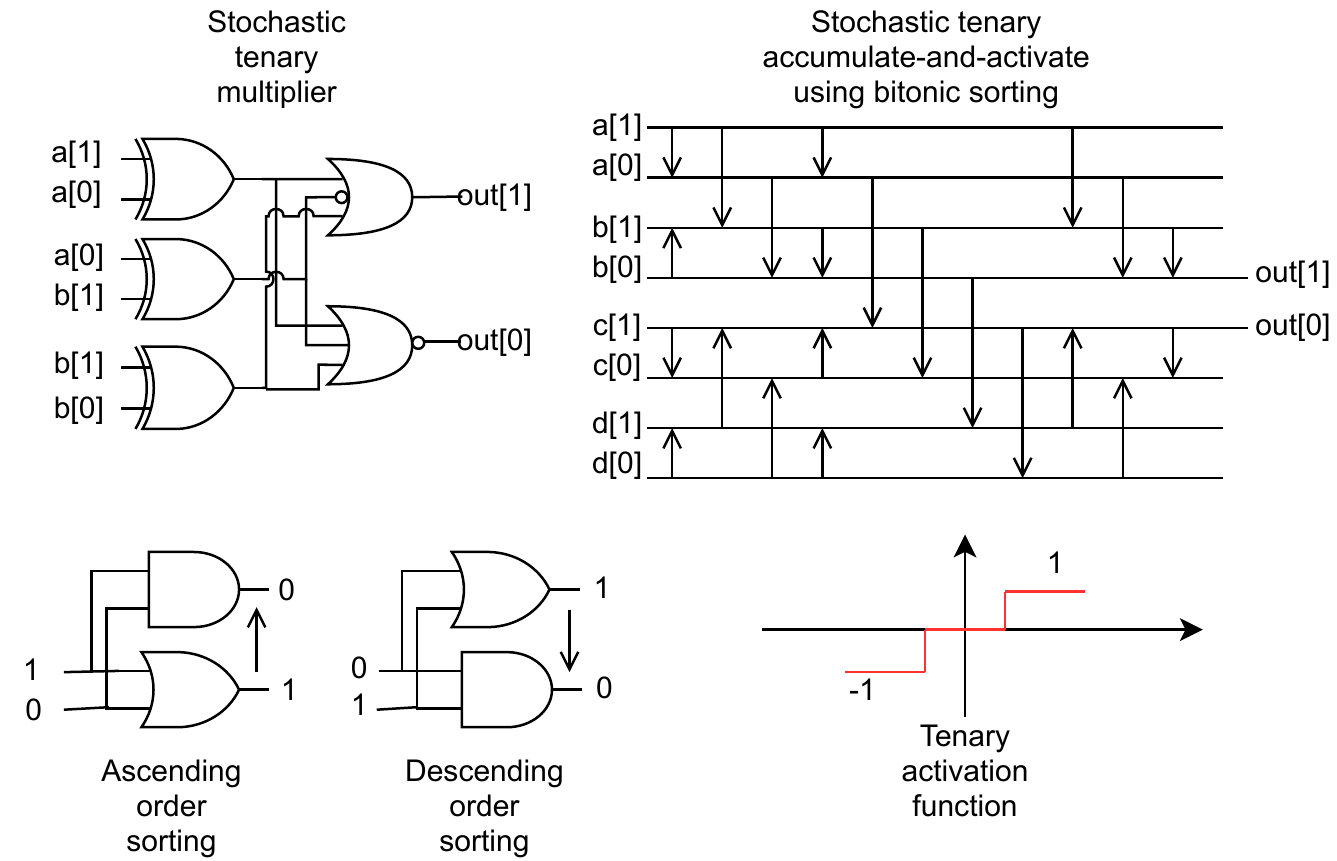}
	\caption{SC circuit implement the function in a TNN. The multiplier is customized to fit the proposed encoding scheme. The accumulate-activate is implemented using a bitonic sorting network circuit.}
	\label{fig:SC_TNN}
\end{figure}

This can be solved by combining SC with low-precision NNs. For example, in \cite{9116390}, a ternary NN (TNN) is implemented by SC using sorting network circuits as the basic processing units while the fault tolerance feature is maintained. Besides, at least 2.8$times$ energy efficiency improvement is obtained compared to its binary counterpart. In TNNs, all the weights and activations are ternarized to $\{0,+1,-1\}$ and they are encoded as $\{10/01,11,00\}$ respectively, with a sequence length of only 2 bits. The complex multiply-accumulate-activate function of each TNN layer (4 input activations and 4 weights) then can be implemented by less than 100 gates. The SC multiplier is tailored for the proposed stochastic encoding scheme and the accumulate-activate is fused and implemented by a bitonic sorting network, as shown in Fig.~\ref{fig:SC_TNN}. The sorting network first place all the 1's on the top and 0's at the bottom. Then the output is decided by the number of 1's and 0's in the input bits. When there are more 1's than 0's, the output is ``11'', indicating a `+1'; when there are more 0's, the output is ``00'', indicating a `-1'; otherwise, the number of 1's and 0's is equal, the output is ``10'', indicting a `0'. This exactly implements the tenary activation function.    

The proposed design is tested on MNIST digit recognition application for its performance, hardware efficiency and fault-tolerance. For the $4\times4\times1$ convolutional layer, the SC-based TNN accelerator can reach 833 TOPS/W, which is 24.5$\times$ of the conventional SC design. The recognition accuracy without any bit-flip error is 97.35\% for the proposed SC design. To evaluate its fault-tolerance, the SRAM read error and multiply-and-add calculation error are considered. When the calculation bit error rate is 10\%, the proposed design maintain a high recognition accuracy of 94\% while that of a binary and conventional SC design is lower than 80\%. An SRAM read error harms the accuracy more than the calculation error, but the fault-tolerance of the SC-based TNN design still outperforms the other two counterparts significantly.   
\subsection{Summary}
A deep neural network model can be error-resilience, therefore, inexact computation results can be tolerated and ``inexact'' circuit design schemes, such as voltage scaling and approximate computing, can be used to improve the performance and energy efficiency. On the other hand, alternative computing paradigms such as SC uses different number encoding system, providing more fault-tolerance, especially against bit-flip error, while dramatically reducing the hardware cost. The long latency used to be a major drawback for SC due to the long sequences required for producing acceptable accuracy. It can be potentially solved by circuit-algorithm codesign with careful optimizations. 
 %fault tolerant circuit design
\section{Cross-layer Fault Tolerance} \label{sec:cross}
In this work, we mainly investigate the fault-tolerant neural network design techniques with a layer-wise manner and the relevant techniques in each layer are illustrated separately in prior sections. However, we still want to emphasize that many cross-layer fault-tolerant approaches have been explored to make best use of the fault-tolerant techniques from different layers for optimized design trade-offs in terms of performance, hardware overhead, and reliability. Hence, we will briefly introduce the cross-layer fault-tolerant approaches with a few typical examples in this section. Jeff Jun Zhang et al. \cite{zhang2018analyzing} \cite{zhang2019fault} proposed to mitigate permanent faults in neural network accelerators with cross-layer optimizations. At architecture layer, a constant bypass is added to each PE in neural network accelerators. On top of the architecture, retraining, a typical model-layer fault-tolerant technique, is applied for each specific fault configuration to recover the model accuracy significantly. The authors in \cite{zhang2018thundervolt} proposed a dynamic per-layer voltage underscaling circuit on top of a classical neural network accelerator such that the accelerator can operate at optimized voltage in each layer of the neural network processing. On top of the circuits, runtime pruning architecture like zero-skip \cite{albericio2016cnvlutin} \cite{7551399} is integrated for higher energy efficiency. Sung Kim et al. \cite{kim2018matic} proposed to combine adaptive neural network training and weight memory voltage scaling to achieve energy-efficient neural network processing. Similar cross-layer optimizations that utilize voltage scaling and fault-aware training or high-level fault correction are also applied in many different scenarios \cite{tu2018rana} \cite{wang2017resilience} \cite{li2019squeezing} \cite{marty2020safe}. In summary, cross-layer fault-tolerant approaches show promising results in generally and it can be expected many of the fault-tolerant techniques surveyed in prior sections can also be potentially combined and optimized for more effective protection against hardware faults. % cross-layer design
\section{Conclusion} \label{sec:conclusion}
In this paper, we reviewed the techniques for fault-tolerant deep learning against perturbations caused by hardware faults in the underlying silicon-based computing engines especially deep learning accelerators. The review is generally conduced in a top-down manner and investigates the fault-tolerant approaches from model layer, architecture layer, and circuit layer respectively. Meanwhile, cross-layer approaches that combine fault-tolerant techniques in multiple layers are also briefly introduced. While fault-tolerant deep learning needs to consider not only the fault tolerance but also many other metrics including performance, accuracy, and hardware overhead at the same time, cross-layer approaches that can leverage advantages of the fault-tolerant techniques from different layers can be potentially beneficial and some of prior work also confirms the great advantages. Nevertheless, cross-layer approaches require synergistic efforts of researchers from AI domain, architectural domain, and reliability domain. 

\section*{Acknowledgment}
The authors would like to thank the support from National Key Research and Development Program of China under Grant No.2020YFB1600201 and National Natural Science Foundation of China (NSFC) under Grant No.62174162, No. 62171313, and No.61902375.

\bibliographystyle{IEEEtran}
\bibliography{ref}

\end{document}